\newcommand{\Iffi}{\Longleftrightarrow}
\newcommand{\bs}{\backslash}
\newcommand{\ket}[1]{| #1 \rangle}
\newcommand{\bra}[1]{\langle #1 |}
\newcommand{\vxv}[1]{\ket{#1}\bra{#1}}
\newcommand{\brkt}[2]{\langle #1 \mid #2 \rangle}
\newcommand{\sqbk}[2]{\bigl| \langle #1 \mid #2 \rangle \bigr|^2}
\documentstyle[12pt]{article}

\title{QUANTUM COMPUTATIONS AND IMAGES RECOGNITION}

\author{Alexander Yu. Vlasov\\
\small Federal Center for Radiology, IRH \\
\small 197101, Mira Street 8, St.-Petersburg, Russia}
\date{23 Sep 1996}
\begin{document}
\maketitle

\begin{abstract}

The using of quantum parallelism \cite{deutsch:turing,deutsch:gates}
is often connected with consideration of quantum system with huge
dimension of space of states. The $n$--qubit register
can be described by complex vector
with $2^n$ components (it belongs to $n$'th tensor
power of qubit spaces). For example, for algorithm of
factorization of numbers \cite{shor:fact} by quantum computer $n$ can
be about a few hundreds for some realistic applications for
cryptography. The applications described further are used some
other properties of quantum systems and they do not demand such huge
number of states.

The term {\it ``images recognition''} is used here for some broad
class of problems. For example, we have a set of some objects
$V_i$ and function of ``{\it likelihood}'':
        $$F(V,W) < F(V,V) = 1; \quad (V \neq W).$$
If we have some ``noisy'' or ``distorted'' image W, we can say that
recognition of $W$ is $V_i$ if $F(W,V_i)$ is near 1 for some $V_i$.

On the other hand, the description of measurement process in quantum
mechanics describes probability of registration as $\Pr = \left|(V,W)\right|^2$, there vector
$V$ describes the system,  $W$ - measurement device, and $(V,W)$ is a
scalar product. So, some kind of ``likelihood'' function is ready to use
in quantum computers.  Of course, the simplest examples are Stern-Gerlach
experiment for electrons and polarizer for photons.

In previous work \cite{vlasov:expsys} as an example of using
quantum computers for some applied problem was chosen simple
expert system.  There are many other similar approaches was found in
different areas later. It is associative memory \cite{kohonen:assocmem},
fuzzy logic \cite{terano:fuzzy}, artificial neural networks,{\it  etc.}\@.
These models are used for further research of the subject.
\end{abstract}

\pagebreak

\section{Introduction}

Richard Feynman \cite{feynman:comp} has compared using of continuous
properties of quantum systems for building of quantum computers with using of
transistors only as two-state systems in electronic digital computer. On the
other hand, {\it analogue computers} use
continuous properties of physical systems for modeling. In the paper is
considered some kind of ``{\it analogue}'' quantum computers those use
properties of quantum systems for building linear models for data analysis%
\footnote{Of course, the {\it quantum parallelism} also uses whole state space
of quantum system}. The digital computers are much more common and convenient
due to {\it universality}, but quantum systems also possess some kind of
universality \cite{deutsch:turing}. There is also some possibility that
the same quantum computer could implement both analogue
modelling discussed further and parallel digital computation.

There are exist few examples of using some models of classical physics for
advanced processing of information --- algorithms based on models of
statistical physics are widely used for data analysis, image recognition
\cite{winkler:image}, learning of artificial neural networks, {\it etc.}\@.

The some examples of using {\it quantum} systems for particular tasks like
{\it image analysis} are discussed further. Linear operators and vector
spaces are not only usual language of quantum mechanics, but they are used
also in many works devoted to such AI problems as associative memory, image
recognition, parallel distributive processing, {\it etc.}\@. Due to such
properties the quantum systems can be used for {\it analogue} computing of
different linear models in data analysis.

The main problem of using of quantum systems is to choose the algorithms of
modeling those are compatible with quantum laws. For example many works
devoted to artificial neural network use noncompact space like {\bf R}$^n$ and some
nonlinear functions together with linear operators. Such models are seem
hardly compatible with language of quantum mechanics. The impossibility of
getting the full information about a quantum system by measurement and the other specific
properties of quantum systems are also must be taken into account. On the other
hand, the more appropriate models we are using for {\it analogue} modeling by
some quantum system, the more elementary and simple system we can use. It is
useful for design of very {\it miniature} and {\it fast} quantum devices.

\section{The spaces of images}

The space of images is usually described as some vector space {\bf V}. For
example, a monochrome digital picture $n \times m$ can be described as a set of
$N = n\, m$ real numbers corresponding to the intensity of light in each
{\it pixel} (picture element), {\it i.e.} some vector ${\bf x} \in {\bf R}^N$
\cite{kohonen:assocmem,winkler:image}.

The {\it hyperspheres}
 $\{{\bf S}^k : {\bf x} \in {\bf R}^{k+1}, \|{\bf x}\| = 1 \}$
and {\it projective spaces} (spaces of rays)
$\{ {\bf R}P\/^k : [{\bf x}] \in {\bf R}^{k+1}_{\bs\{0\}}/{\bf R}_{\bs\{0\}};\:
[x_0{:}x_1{:}\ldots{:}x_k] = [\lambda x_0{:}\lambda x_1{:}\ldots{:}\lambda x_k];
{\bf x} \neq {\bf 0}, \lambda \neq 0 \}$
are also can be used.

For example, in case of monochrome picture we can multiply all intensities on
the same nonzero positive value and the picture does not change. Due to such
invariance we can use only vectors with unit length and space of images is
subspace of the sphere ${\bf S}^{N-1}$.

An {\it image recognition} can be described as following task. There are $m$
{\it known} images: ${\bf v}^{(1)} \ldots \ {\bf v}^{(m)} \in {\bf V}$.
A system should ``{\it recognize}'' any image ${\bf v}^{(i)}$ by his {\it noisy}
or {\it incomplete} version {\bf w}. If the {\bf w} is not precisely equal to
some ${\bf v}^{(i)}$ we can use a measure on {\bf V} to choose ${\bf v}^{(i)}$
that more ``close'' to {\bf w}. Often such a measure is {\it Euclidean
distance} on ${\bf R}^n$:
\begin{equation}
 \| \vec{w} - \vec{v} \|_{R^n} \equiv \sqrt{\sum_{i=1}^n (w_i - v_i)^2}
\label{eq:eucnorm}
\end{equation}

In case of ${\bf S}^{n-1}$ the {\it cosine} of angle between two vectors with
unit length is:
\begin{equation}
 \cos(\angle_{\vec{w},\vec{v}}) = \sum_{i=1}^n w_i v_i \equiv
 \left( \vec{w},\vec{v} \right)
\label{eq:sphnorm}
\end{equation}

The $\left({\bf w, v} \right)$ is scalar product and for vectors with unit
norm:
\begin{equation}
 \left(\vec{w},\vec{v} \right) = 1 \Iffi  \vec{w} = \vec{v}
\label{eq:maxproj}
\end{equation}

The property (\ref{eq:maxproj}) is used in some approaches for image
recognition. For space of images like ${\bf S}^k$ criterion
$ 1 - \varepsilon < ({\bf w, v}^{(i)}) \le 1 $ can be used.
It should be mentioned that the measures are good for models of errors like
addition of some random noise or lack of some areas in picture. The same
image after {\it moving} or {\it rotation} may have very low correlation with
initial image from point of view of above mentioned criterion.

It should be mentioned that the more {\it homogeneously} distribution of images in
the space the better. For example we can subtract {\it half of average intensity}
from any points of the monochrome picture before normalization

\begin{equation}
y_i = x_i - \frac{1}{2 N} \sum_{i=1}^N{x_i}, \quad
\vec{z} = \frac{\vec{y}}{\| \vec{y} \|}
\label{eq:averz}
\end{equation}

In case of homogeneous distribution on the ${\bf S}^N$ the scalar product
of two random vectors is:
\begin{equation}
 \left(\vec{w},\vec{v} \right) \sim N^{-1/2}
 \stackrel{N \to \infty}{\longrightarrow} 0
\label{eq:scalz}
\end{equation}

\medskip

If we are going to use quantum systems for analogue modeling, then using of
approaches with ${\bf S}^k$ or ${\bf R}P\/^k$ is especially justified.

\section{The Hilbert spaces}

The space of states of quantum system is described by {\it Hilbert space}
${\cal H}$, {\it i.e.} complex vector space with Hermitian scalar product:
\begin{equation}
 \left({\bf a} , {\bf b} \right) = \sum_{i=1} a_i \overline{b_i} \ , \quad
 \| {\bf a} \| = \sqrt{\left( {\bf a}, {\bf a} \right)}
\label{eq:hermnorm}
\end{equation}
In physics there are notations $\ket{a}$ for the vector ${\bf a}$ and
$\bra{a}$ for co-vector ${\bf a}^*$, and
\begin{equation}
\brkt{b}{a} = {\bf b}^* {\bf a} = ({\bf a},{\bf b}).
\label{eq:quantnorm}
\end{equation}

The vectors $\ket{\psi}$ and $\lambda \ket{\psi}$ for any
$\lambda\in{\bf C} - \{0\}$ are describe the same physical state. The states
of quantum systems are rays in complex vector space {\it i.e.} points in {\it
complex projective space} ${\bf C}P^n$ (or ${\bf C}P^\infty$). Due to projectivity,
we can consider only states with unit norm $\|\psi\| = 1$. It is hypersphere
${\bf S}^{2n+1} \subset {\bf R}^{2n} \simeq {\bf C}^n$. The vectors have
property maximum projection (eq. \ref{eq:maxproj}) for scalar product
(eq. \ref{eq:hermnorm}), but for the same physical state $\ket{\psi}$ and
$\ket{\psi'} = e^{i\varphi} \ket{\psi}$:
$\brkt{\psi}{\psi'} = e^{i\varphi} \neq 1$.

The {\t different} quantum states correspond to space of rays
${\bf C}P^n \simeq {\bf S}^{2n+1}/{\bf S}^1$. The analogue of property
(\ref{eq:maxproj}) for space of states of quantum system is:
\begin{equation}
 \Pr(\chi \to \psi) \equiv \sqbk{\psi}{\chi} = 1
 \Iffi  \ket{\chi} = \lambda \ket{\psi} \Iffi [\psi] = [\chi]
\label{eq:maxprob}
\end{equation}
Where $[\psi], [\chi] \in  P({\cal H})$ are complex {\it rays} in Hilbert
space those correspond to states of quantum system.

\section{Quantum systems}

Let us consider physical meaning of the algebraic formulae above.  The
$\Pr(\chi \to \psi)$ is probability of registration of system $\ket{\chi}$ by
the measurement device that works as a filter for state $\ket{\psi}$ \cite{feynman:flph3}.
It is some kind of binary {\sf Yes / No} consideration. We are ``asking'', does
the system in state $\ket{\psi}$, and for quantum system $\ket{\chi}$ we have
``answer'' {\sf Yes} with probability $\Pr(\chi \to \psi) = \sqbk{\psi}{\chi}$.
We can see, that $\Pr = 1$ {\it if and only if\/} $\ket{\chi}$ and
$\ket{\psi}$ is the same physical state {\it i.e.} the same point on complex
projective space.
There is second model of measurement that also will be useful further. If
we have some orthogonal basic states $\ket{\psi_i}$ and quantum system $\ket{\chi}$
\begin{equation}
 \|\chi\| = 1, \quad \brkt{\psi_i}{\psi_j} = \delta_{ij}, \quad
 \ket{\chi} = {\sum_{i=1}^n a_i \ket{\psi_i}}, \; a_i = \brkt{\chi}{\psi_i} .
\label{eq:qubasis}
\end{equation}
then we can get one of $\ket{\psi_i}$ due to measurement of the $\ket{\chi}$ with
probability $\Pr_i = ( a_i )^2$, {\it i.e.} have got one of {\it n}
versions of outcome instead of two in the previous example. But $\ket{\psi_i}$
in this case must be {\it orthogonal}. This simple quantum mechanical
idea has emphasized here for further discussion about using of {\it orthogonal}
set of images.

\medskip

The consideration of particular problems of data analysis that can be related
with the mentioned approach are followed. Further will be
used examples of data analysis with compact space of images like
${\bf R}P^n,\: {\bf S}^n$ that was developed by different authors.
%(${\bf C}P^n = {\bf S}^{2n+1}/{\bf S}^1$)
Such kind of models can be considered as a good starting point due to analogy
between conditions of maximum of (\ref{eq:maxproj}) for real spaces and
(\ref{eq:maxprob}) for complex space of state. It should be mentioned also
that ${\bf R}P^n \subset {\bf C}P^n$ and so an $\upsilon \in {\bf R}P^n$ can
be treated as some formally {\it possible} state of quantum system
{\it i.e.} it is not contradict with laws of quantum mechanics.

\section{The quantum systems and analogue image recognition}

The using of classical device for analogue calculation of the expressions like
(eq.\ref{eq:sphnorm}) for scalar products \cite{kohonen:assocmem} was more
justified before the {\it digital computers} have become fast and cheap
enough. The quantum computers could make the such approach useful again.

For realization of the above mentioned algorithms of image recognition
we should add to usual unitary operations of quantum computer a new
one. The operation is similar to {\it measurement} in usual description of
macroscopic experiment \cite{feynman:flph3}. It can be described as
{\it transition} from state $\ket{\psi}$ to state $\ket{\chi}$ with probability
$\sqbk{\chi}{\psi}$. In description of quantum computers
the related effects are usually considered as undesirable sources of errors%
\footnote{The same expression $\sqbk{\chi}{\psi}$ is called
{\it fidelity} of quantum channel \cite{schumacher:qubit,bennet:review}}.
Here is discussed an useful application of the effects.

\medskip

Let us consider some method of {\it input} of the images data
to quantum computer as smooth map from the space of images to the
space of states of quantum system: ${\cal I} : {\bf V} \to {\cal V \subset H}$.
\footnote{Here {\it smooth} means {\it continuous} from respect of some
norms on {\bf V} and ${\cal H}$.}

In this case simple quantum {\it read-only memory} (q-{\sf ROM}) for {\it one}
image can be considered as a filter $\vxv{\psi_{image}}$ that receive some
$\ket{\psi}$ as an input and produce $\ket{\psi_{image}}$ as the output
with probability $\Pr = \sqbk{\psi_{image}}{\psi}$ that is
equal to {\it one} if $\ket{\psi} = \ket{\psi_{image}}$.

If distribution of the inputs is {\it homogeneous} and dimension of state space
$N$ is big enough, then scalar product is ``close'' to {\it zero}  for some
arbitrary input due to (eq.\ \ref{eq:scalz}). Because of this property the
probability of recognition is very small for arbitrary images ($\Pr \sim
1/N$). On the other hand, for images that differ from true image due to some
small errors the recognition is near to {\it one}.

\begin{figure}[htbp]
\begin{center}
\unitlength=1.00mm
\special{em:linewidth 0.4pt}
\linethickness{0.4pt}
\begin{picture}(92.00,52.00)
\put(48.00,8.00){\line(1,1){4.00}}
\put(48.00,18.00){\line(1,1){4.00}}
\put(48.00,28.00){\line(1,1){4.00}}
\put(48.00,38.00){\line(1,1){4.00}}
\put(48.00,48.00){\line(1,1){4.00}}
\put(15.00,10.00){\vector(1,0){15.00}}
\put(50.00,10.00){\vector(1,0){15.00}}
\put(50.00,20.00){\vector(1,0){15.00}}
\put(50.00,30.00){\vector(1,0){15.00}}
\put(50.00,40.00){\vector(1,0){15.00}}
\put(50.00,50.00){\vector(1,0){15.00}}
\put(50.00,10.00){\vector(0,1){5.00}}
\put(50.00,15.00){\vector(0,1){10.00}}
\put(50.00,25.00){\vector(0,1){10.00}}
\put(50.00,35.00){\vector(0,1){10.00}}
\put(80.00,6.00){\framebox(12.00,8.00)[cc]{$\vxv{\iota_1}$}}
\put(80.00,16.00){\framebox(12.00,8.00)[cc]{$\vxv{\iota_2}$}}
\put(80.00,26.00){\framebox(12.00,8.00)[cc]{$\vxv{\iota_3}$}}
\put(80.00,36.00){\framebox(12.00,8.00)[cc]{$\vxv{\iota_4}$}}
\put(80.00,46.00){\framebox(12.00,8.00)[cc]{$\vxv{\iota_5}$}}
\put(30.00,10.00){\line(1,0){20.00}}
\put(65.00,10.00){\line(1,0){15.00}}
\put(65.00,20.00){\line(1,0){15.00}}
\put(65.00,30.00){\line(1,0){15.00}}
\put(65.00,40.00){\line(1,0){15.00}}
\put(65.00,50.00){\line(1,0){15.00}}
\put(50.00,45.00){\line(0,1){5.00}}
\put(3.00,6.00){\framebox(12.00,8.00)[cc]{\it input}}
\put(20.00,12.00){\makebox(0,0)[lb]{$\ket{\chi_{input}}$}}
\end{picture}
\end{center}
\caption{Splitting of dense beam}
\label{fig:split}
\end{figure}
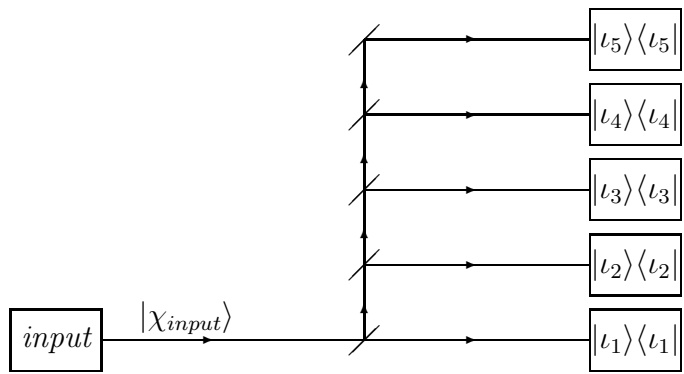

If we have a {\it classical} {\it input} like monochrome picture then there
is possibility of repeatedly preparing of the same $\ket{\psi(input)}$
to satisfy desirable statistical criteria for any number of different
q-{\sf ROM}s with different images that would be enough to find an image
with maximum probability $\Pr_i = \sqbk{\psi(input)}{\psi(image_i)}$.
On (Fig. \ref{fig:split}) is shown source of quantum systems controlled by
some {\it input}. The $\ket{\chi_{input}}$ can be considered for example as
molecular beam that is split by some partially transparent passive ``mirrors''
to $n$ arms with filters $\iota_1 \ldots \iota_n$. If the intensities before
the filters in each arm are the same ($I_0/n$) then intensities of the beams
after filters are $I_k = \sqbk{\iota_k}{\chi_{input}}I_0/n$. The channel that
correspond to restored image has maximum intensity.

The more effective procedure can be used in case of {\it orthogonal} images.
If the images is not quite orthogonal, but they have scalar products
near zero (see eq.\ \ref{eq:scalz}) it is possible to use standard method
of ortogonalization of vectors and use new orthogonal set of corrected
images. The quality of such methods for real problems of images recognition is
discussed in \cite{kohonen:assocmem}.

Let us suppose that we have such a set of orthogonal images. Then it is
possible to make measurement that recognize any of $k$ images with
$\Pr = 1$. The such measurement is described by (eq. \ref{eq:qubasis}) if
all images correspond to some basic vectors $\ket{\psi_i}$ in
(eq. \ref{eq:qubasis}). For example it may be first $k$ vectors
$\ket{\psi_1} \ldots \ket{\psi_k},\, k \ll n$.

\begin{figure}[htbp]
\begin{center}
\unitlength=1.00mm
\special{em:linewidth 0.4pt}
\linethickness{0.4pt}
\begin{picture}(120.00,36.00)
\put(5.00,20.00){\circle*{2.00}}
\put(10.00,18.00){\line(0,-1){13.00}}
\put(10.00,5.00){\line(-1,0){10.00}}
\put(0.00,5.00){\line(0,1){30.00}}
\put(0.00,35.00){\line(1,0){10.00}}
\put(10.00,35.00){\line(0,-1){13.00}}
\put(5.00,20.00){\vector(1,0){45.00}}
\put(20.00,20.00){\vector(4,1){20.00}}
\put(20.00,20.00){\vector(4,-1){20.00}}
\put(20.00,20.00){\vector(2,-1){20.00}}
\put(20.00,20.00){\vector(2,1){20.00}}
\put(40.00,22.00){\rule{1.00\unitlength}{14.00\unitlength}}
\put(40.00,4.00){\rule{1.00\unitlength}{14.00\unitlength}}
\put(50.00,10.00){\framebox(20.00,19.00)[cc]{$H_{Image}$}}
\put(100.00,20.00){\vector(4,1){20.00}}
\put(100.00,20.00){\vector(4,-1){20.00}}
\put(100.00,20.00){\vector(2,-1){20.00}}
\put(100.00,20.00){\vector(2,1){20.00}}
\put(42.00,22.00){\makebox(0,0)[lb]{$| \psi_0 \rangle $}}
\put(71.00,22.00){\makebox(0,0)[lb]{$U_{Image} | \psi_0 \rangle $}}
\put(70.00,20.00){\vector(1,0){50.00}}
\put(20.00,25.00){\framebox(8.00,10.00)[cc]{\bf S}}
\put(20.00,5.00){\framebox(8.00,10.00)[cc]{\bf N}}
\put(101.00,25.00){\framebox(8.00,10.00)[cc]{\bf S}}
\put(101.00,5.00){\framebox(8.00,10.00)[cc]{\bf N}}
\put(90.00,5.00){\dashbox{2.00}(8.00,30.00)[ct]{\shortstack{\\$U_r$}}}
\end{picture}
\end{center}
\caption{Recognizing of orthogonal images}
\label{fig:orthim}
\end{figure}
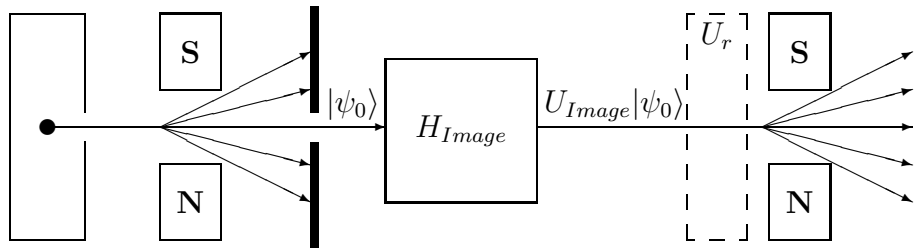

The simple example is shown on (Fig. \ref{fig:orthim}). Here the $U_r$ is unitary operator
that rotate an orthogonal set of images to orthogonal set of basic vectors
of measurement device. By using different $U_r$ it is possible
to make q-{\sf ROM} for any $n$ orthogonal vectors and first $k \ll n$ can be
good approximation for set of images due to conditions like (eq. \ref{eq:scalz}).

\section*{Acknowledgments}
The author is grateful to Roman Zapatrin, Sergey Krasnikov, {\it etc.} for
fruitful discussion about the paper, to Andrew Grib and other participants of
seminar in Friedman Laboratory for Theoretical physics for numerous and
interesting talks, to Andrew Borodinov for many references and discussions
about the such modern AI areas as artificial neural networks, parallel
distributive processing, fuzzy logic, associative memory, {\it etc.}. Many
thanks to Tetsuro Nishino, Peter Shor, Artur Ekert, and Seth Lloyd for sending
copies of articles and preprints.

%\pagebreak

\end{document}